\documentclass[prl,showpacs,amsmath,amssymb,twocolumn,superscriptaddress,raggedbottom,floatfix]{revtex4}
\usepackage{hyperref}
\usepackage{amsmath}
\usepackage{amssymb}
\usepackage{bm}
\usepackage{amsthm}
\usepackage{amsfonts}
\usepackage{latexsym}
\usepackage{graphicx}
\begin{document}

\title{Strain-controlled band engineering and self-doping in ultrathin LaNiO$_3$ films}

\author{E.~J. Moon}
\email{exm014@uark.edu}
\affiliation{Physics Department, University of Arkansas, Fayetteville, AR 72701, USA}

\author{J.~M. Rondinelli}
\affiliation{Department\! of\! Materials\! Science\! \& Engineering,\!
Drexel\! University,\! Philadelphia,\! PA\! 19104,\! USA}

\author{N.~Prasai}
\affiliation{Physics Department, University of Miami, Coral Gables, FL 33124, USA}

\author{B.~A. Gray}
\affiliation{Physics Department, University of Arkansas, Fayetteville, AR 72701, USA}

\author{M.~Kareev}
\affiliation{Physics Department, University of Arkansas, Fayetteville, AR 72701, USA}

\author{J.~Chakhalian}
\affiliation{Physics Department, University of Arkansas, Fayetteville, AR 72701, USA}

\author{J.~L.~Cohn}
\affiliation{Physics Department, University of Miami, Coral Gables, FL 33124, USA}

\begin{abstract}
We report on a systematic study of the temperature-dependent Hall coefficient and thermoelectric power in ultra-thin metallic LaNiO$_{3}$ films that reveal a strain-induced, self-doping carrier transition that is inaccessible in the bulk. As the film strain varies from compressive to tensile at \emph{fixed} composition and stoichiometry, the transport coefficients evolve in a manner strikingly similar to those of bulk hole-doped superconducting cuprates with \emph{varying} doping level. Density functional calculations reveal that the strain-induced changes in the transport properties are due to self-doping in the low-energy electronic band structure. The results imply that thin-film epitaxy can serve as a new means to achieve hole-doping in other (negative) charge-transfer gap transition metal oxides without resorting to chemical substitution.
\end{abstract}


\maketitle

\emph{Introduction.---}%
Atomic-level control over the deposition of transition metal oxide (TMO) films and multilayers has generated considerable interest in harnessing two-dimensional (2D) confinement and misfit-strain to tailor the carrier concentration of correlated electron systems for novel electronic applications\cite{Mannhart/Schlom:2010}. An important research avenue toward this goal is to establish to what extent confinement and strain mimic chemical doping---the principal means for accessing metal-insulator (MI), superconducting and colossal magnetoresistive transitions in perovskite titanates \cite{Imada/Fujimori/Tokura:1998}, cuprates, \cite{Pickett:1989} and manganites \cite{Salamon/Jaime:2001}. Recently, electrostatic hole-doping has been demonstrated in vanadate quantum wells \cite{Higuchi/Hwang:2009} and manganite heterostructures \cite{Hikita/Hwang:2009,Molegraaf/Ahn/Triscone:2009,Burton/Tsymbal:2011,Yajima/Hwang:2011} using interfacial dipole engineering. Here we demonstrate charge-carrier doping in ultra-thin films of the metallic oxide LaNiO$_3$ (LNO) at \textit{fixed stoichiometry} through substrate/film lattice misfit strain alone, without real-space charge transfer (CT) at the heterointerface. The strain-induced self-doping is accomplished through a redistribution of charge among the low-energy O~2$p$ bands and Ni 3$d$~states at the Fermi level.

We choose to study the self-doping effect in the correlated perovskite metal LNO because it has a low-energy electronic structure derived from two bands with $A_{1g}$-symmetry ($d_{z^2}$ and $d_{x^2-y^2}$) at quarter-filling \cite{Rajeev_et_al:1991, Sreedhar/Spalek_et_al:1992} and is classified as a CT oxide \cite{Zaanen/Sawatzky/Allen:1984}. Recent epitaxial film studies demonstrate that biaxial strain-induced lattice distortions and orbital-level energy splitting determine the transport properties \cite{Dobin/Goldman_etal:2003, May/Rondinelli:2010, Boris/Keimer_et_al:2011, Chakhalian/Rondinelli:2011}, orbital polarization \cite{Han/Millis_etal:2010, Benckiser/Keimer:2011} and strength of electron-electron correlations \cite{Stewart/Haule/Chakhalian_etal:2011,Oullette/Allen_etal:2010}, indicating that the near-Fermi surface (FS) electronic structure is highly sensitive to misfit strain.

Furthermore, LNO is considered a Cu-free analogue to the cuprates \cite{Anisimov/Bukhvalov/Rice:1999,Chaloupka/Khaliullin:2008,Hansmann/Held_et_al:2009}, as it is also a covalent strongly correlated material with a partial $d_{x^2-y^2}$ band filling that governs transport behavior. However, LNO differs in important respects from cuprates: it is neither 2D nor a doped Mott insulator in bulk. There is no nickelate equivalent to the antiferromagnetic, $S=\frac{1}{2}$ insulating cuprate parent compound \cite{Anisimov/Bukhvalov/Rice:1999}. Nonetheless, our Hall and thermoelectric power (TEP) studies, over a range of dopings inaccessible in bulk, reveal a surprising commonality with doped cuprates not previously appreciated. The implication is that substrate-induced misfit strain in nickelate films reproduces the effects of direct cation substitution (charge-carrier doping) found in cuprates. Density functional calculations performed within the local spin density approximation plus  Hubbard $U$ (LSDA$+U$) method corroborate the \textit{misfit strain-induced self-doping} picture, indicating that epitaxially strained LNO films are proximal to a self-doped MI-transition as are rare-earth substituted nickelates \cite{Mizokawa/Khomskii/Sawatzky:2000}.

\emph{Growth and Characterization.---}
Epitaxial 10~unit cell (u.c.) ultra-thin films of LNO were grown on YAlO$_3$ (YAO), SrLaAlO$_{4}$ (SLAO), LaAlO$_{3}$ (LAO), SrLaGaO$_{4}$ (SLGO), (LaAlO$_{3}$)$_{0.3}$(Sr$_{2}$AlTaO$_{6}$)$_{0.7}$ (LSAT), SrTiO$_{3}$ (STO), and GdScO$_{3}$ (GSO) (001)-oriented substrates by pulsed laser deposition following the procedures in Ref.~\cite{Kareev/Chakhalian:2011}. A physical property measurement system was employed to measure the dc Hall coefficient ($R_{H}$) and resistivity ($\rho$) of the films with fields up to $B=3$~T applied perpendicular to the film plane. The TEP was measured with a steady-state technique using a fine-wire chromel-constantan thermocouple and gold leads.

\begin{figure}
  \centering
  \includegraphics[width=\columnwidth]{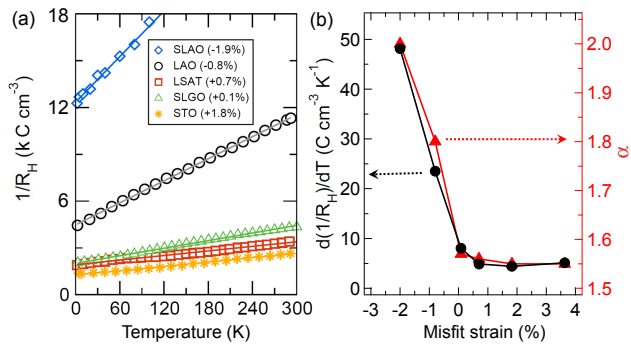}
  \caption{(Color online) (\emph{a}) Temperature dependent inverse Hall coefficient $R_H^{-1}$, (\emph{b}) temperature derivative of $R_H^{-1}$ (left ordinate) and $\cot\theta_H$ power-law temperature exponent $\alpha$ (right ordinate) for LNO films \emph{vs.} misfit strain.}
  \label{fig:fig1}
\end{figure}

\begin{figure*}
  \centering
  \includegraphics[width=2\columnwidth]{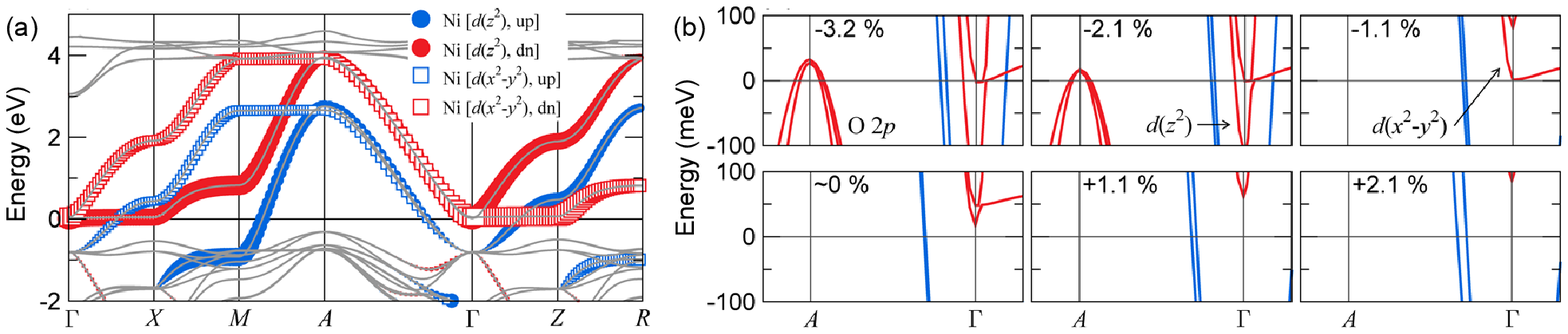}
  \caption{(Color online) (\emph{a}) Electronic band structure plot for LNO at $\sim 0\%$ strain. The majority-spin (blue, dark grey) and minority spin (red, light grey) are highlighted for the spin-polarized (up or down) Ni $d_{z^2}$ and $d_{x^2-y^2}$ orbitals. The symbols' size correspond to the magnitude of the Bloch states projected onto each atomic orbital. (\emph{b}) Low-energy electronic structure for tetragonal $P4/mmm$ LNO about the $\Gamma$- and $A$-points at different biaxial strain states.}
  \label{fig:e-structure}
\end{figure*}

\emph{Transport.---}%
Figure~\ref{fig:fig1}(\emph{a}) shows the temperature-dependent inverse Hall coefficient $R_H^{-1}(T)\equiv n_He$ for LNO films under a range of strains. As seen $R_H^{-1}$ is linear in $T$ and positive for all $T$ and systematically decreases in magnitude with increasing tensile (positive) strain. Note, in simple one-band metals $R_H$ is independent of $T$. The asymmetric strain dependence of the Hall data is revealed when we plot the slope $dR_{H}^{-1}/dT$ and power-law ($T^\alpha$) temperature exponent $\alpha$ of the inverse Hall angle, $\cot\theta_H=\rho/(R_HB)$, against strain [Fig.~\ref{fig:fig1}(b)]. As shown, both the effective carrier density and scattering are highly sensitive to compressive strains, while showing little (or no) variation under tensile strains.

The observed $T$-dependent Hall data bear a striking resemblance to those of cuprates, where $R_H^{-1}$ is also approximately linear in $T$ over a broad temperature range. The $R_H^{-1}\propto T$ and $\cot\theta_H\propto T^2$ dependencies for compressively strained films [Fig.~\ref{fig:fig1}(\emph{b})], as for underdoped cuprates, can be reproduced within the single-band Boltzmann theory by a Fermi-liquid-like $T^2$ scattering rate that varies sharply about the FS \cite{Hussey:2008, Hussey:2003, Ioffe/Millis:1998}. Within this picture, these characteristic $T$ dependencies directly reflect the scattering anisotropy, and their weakening in LNO films with increasing tensile strain signals a decrease in the effective scattering anisotropy. In close similarity, a decrease in the power-law exponent $\alpha$ from 2 to 1.65 with increasing hole doping for Bi-based cuprates \cite{Ando/Murayama:1999,Konstantinovi:2000} has been attributed to increased structural disorder that tends to smear out scattering anisotropy on the FS \cite{Hussey:2003, Hussey:2008}. Alternatively, a strong $T$-variation of $R_H^{-1}$ is possible when electron and hole contributions (in a two-band model) possess different $T$-dependent mobilities. However, recent photoemission observations of a substantial mass enhancement for the electron pocket in LNO \cite{Eguchi/Shin_et_al:2009}, but not for the hole pocket, suggest that the respective carrier conductivities obey $\sigma_h\gg\sigma_e$, and hence a single band picture may indeed be appropriate. The striking similarities between the strain-dependent Hall data in LNO films and those of bulk hole-doped cuprates suggest that the asymmetric strain-dependent scattering has its origin in a changing symmetry and/or character of electronic states at the Fermi level ($E_F$).

\emph{Electronic structure.---}
To better understand the unusual strain-dependent Hall data, we carry out first-principles density functional calculations on LNO films using the local spin density plus Hubbard $U$ approximation \cite{footnote1}. Figure~\ref{fig:e-structure}(\emph{a}) depicts the electronic band structure of LNO at $\sim0$~\% strain using the ``fat-bands'' method \cite{Jepsen/Andersen:1995}, which allows us to distinguish the different orbital character of the Bloch states near the Fermi level. Consistent with the ionic picture of low-spin Ni$^{3+}$, the states at $E_F$ are primarily of Ni $d_{z^2}$ and $d_{x^2-y^2}$ symmetry, with partial occupation of majority-spin (blue) Ni-states. The minority-spin (red) Ni states are exchange split by $\sim0.80$~eV to higher energy, making LNO under zero strain a half-metal.

In addition to the Ni $e_g$-bands at $E_F$, there are low-lying oxygen 2$p$ bands located less than 0.50~eV below the Fermi level. The fat-band analysis shows that these states at the $A$-point, $k=(\frac{1}{2},\frac{1}{2},\frac{1}{2})$, are completely of oxygen character---no hybridization with the Ni $d$-states occurs. Because these states are purely oxygen-derived, they could act as a source (sink) for holes (electrons) to the nearby Ni $d$-states if they become partially occupied \cite{Korotin/Sawatzky:1998,Khomskii:1997}. This electronic configuration, with O 2$p$ states close in energy to the occupied Ni $d$-manifold, classifies LNO as a negative CT-gap compound susceptible to self-doping \cite{Mizokawa/Khomskii/Sawatzky:2000}.

Consider the changes in low-energy electronic structure about the high-symmetry $\Gamma$- and $A$-points with misfit strain [Fig.~\ref{fig:e-structure}(\emph{b})]. On going from compressive to tensile strain, an electronic transition from fully metallic -- both spin channels are occupied at $E_F$ -- to a half-metallic state occurs. In other words,  The metallic state present under compressive strain requires a change in the hole population at the Fermi level in order to keep the chemical potential fixed \cite{footnote2}. Indeed, the O~2$p$ states, fully occupied at zero strain, are partially empty at the $A$-point under compressive strain: electrons are transferred from the minority-spin O~2$p$ states to the minority spin Ni-states.
Specifically, the $d_{z^2}$ orbital at $\Gamma$ is partially occupied with holes [cf.\ -3.1~\% and -2.1~\% strain panels in Fig.~\ref{fig:e-structure}(\emph{b})]. Consistent with this CT to the minority-spin manifold, there is a reduction in the Ni-spin moment under compressive strain (0.4~$\mu_B$/Ni) compared to that at zero strain (0.75~$\mu_B$/Ni). In other words, biaxial compressive strain induces a self-doping effect on the FS at fixed composition. As a result, the relative volume of the electron (hole) pocket increases (decreases) as the strain goes from compressive to tensile, qualitatively consistent with the trend in $R_H^{-1}$. While the precise band occupancies as a function of strain depend on the value of $U$, the self-doping and evolution of the electronic structure are robust features of the calculations.

\begin{figure}[b]
  \centering
  \includegraphics[width=\columnwidth]{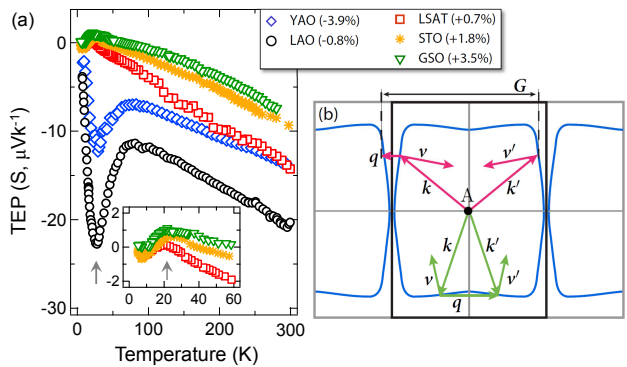}
  \caption{(Color online) (\emph{a}) $T$-dependent TEP for
  strained LNO films. (\emph{b}) Scattering geometry on the majority-spin FS sheet centered at the $A$-point for -2.1\% strain: normal (lower, green) and Umklapp (upper, pink) processes are shown (the latter involving reciprocal lattice vector $\textbf{\em G}$).}
  \label{fig:tep}
\end{figure}

\emph{Self-doping.---}%
The temperature-dependent TEP ($S$) for strained LNO films further confirms the proposed self-doping picture [Fig.~\ref{fig:tep}(\emph{a})]. The TEP for all films at $T\geq 120$~K is accurately described by a linear-$T$ form ($A+BT$), with $B=-$0.041$\pm$ 0.004 $\mu$V/K$^{2}$ approximately independent of strain, and with a constant offset $A$ that increases with tensile strain (the film grown on YAO is an exception \cite{footnoteYAO}). If the linear-$T$ term is interpreted as diffusion-TEP, the value of $B$ allows the calculation of the Fermi energy using the Mott expression $S = \pi^{2}k_{B}^2T/(3eE_{F})$ under the assumption of a linear energy dependence of the conductivity at the Fermi level \cite{footnote3}. We find $E_F\approx 0.6$~eV, in good agreement with recent photoemission data on LNO \cite{Eguchi/Shin_et_al:2009}.

These systematics of the TEP are also strikingly similar to those of the normal-state TEP of bulk, superconducting cuprates. For cuprates, the constant term $A$ increases with {\it decreasing} hole doping of the CuO$_2$ planes, achieved via cationic substitution \cite{Obertelli:1992,Honma/Hor:2008}. This hole-doping behavior matches the trend that we observe in our stoichiometric films with strain: the self-doping decreases ($A$ increases) as the strain state goes from compressive to tensile. Remarkably, no change in composition is needed in the LNO films to reproduce the cuprate-like transport.

A sharp peak in the TEP near $T=25$~K (arrowed) also correlates with strain [Fig.~\ref{fig:tep}(\emph{a})] \cite{Greene:98}. It is large and negative for films under compressive strain and small and positive for those under tension [Fig.~\ref{fig:tep}(\emph{a}), inset]. The magnetic-field independence (to 9~T) of this feature (data not shown) argues against its magnetic origin.

Low-$T$ TEP peaks in metals are typically attributed to phonon drag. To reproduce the behavior of the LNO films, a negative phonon-drag TEP contribution that becomes positive and weakens in the misfit strain progression from compressive to tensile is needed. Because strain affects the electronic structure near the $E_F$ [Fig.~\ref{fig:e-structure}(b)], the scattering geometry relevant to phonon-drag TEP ($S_g$) should also be significantly altered. For LNO under $-2.1$\% strain, the majority-spin FS sheet centered at the $A$-point nearly touches the Brillouin zone (BZ) boundary [Fig.~\ref{fig:tep}(\emph{b})]. Superimposed on the FS are the normal (lower, green) and Umklapp (upper, pink) scattering processes ($\textbf{k}\to {\textbf k^{\prime}}$) involving phonon wavevectors ${\bf q}$ that cross unoccupied regions of the BZ. Both yield $S_g\propto -\Delta {\textbf v}\cdot {\textbf q} < 0$ (${\textbf v}$ is the Fermi velocity). We anticipate the Umklapp contributions to be particularly strong given the flatness of the FS near the BZ boundary. As this FS sheet shrinks and moves away from the BZ boundary with increasing tensile strain [Fig.~\ref{fig:e-structure}(\emph{b})], the Umklapp contribution will be suppressed as the low-$T$ phonon population with increasingly large spanning vector ${\textbf q}$ is sharply depleted. The disappearance of these Umklapp scattering contributions to phonon drag is consistent with the disappearance of the sharp negative peak in the TEP near 25~K that occurs under tensile strain [Fig.~\ref{fig:tep}(\emph{a}), inset], further supporting the concept of strain-induced self-doping.

\emph{Carrier doping of CT oxides.---}
Our calculations indicate that the ground-state electronic structure and oxygen hole density are directly modulated by biaxial strain: as the strain changes from compressive to tensile, the O~2$p$ states shift to lower binding energy. When strain approaches zero, the oxygen levels become fully occupied and self-doping is quenched. This shift in oxygen-band filling mirrors the change in occupation of the minority-spin Ni $e_g$ states at $\Gamma$. The orbital-level splitting between the $d_{z^2}$ and $d_{x^2-y^2}$ states decreases as biaxial strain approaches zero; it increases with larger tensile strains, but both minority-spin Ni $d$-states remain unoccupied, producing the half metallic ground state. Thus strain, which directly alters the crystal field, is responsible for the self-doping. 

A small or negative CT gap, occurring when the ligand $p$-levels lie close in energy to or overlap the $d$-levels (upper Hubbard band), is a key ingredient to the strain-controlled self-doping of LNO films. Other TMO compounds, particularly perovskites involving late $3d$ and $4d$ transition metals, likely fall into this category \cite{Korotin/Sawatzky:1998, Khomskii:1997, Khomskii/Sawatzky:1997}. It is anticipated that the self-doping may be manipulated through strain in some of these compounds in a manner similar to that demonstrated here, \textit{i.e.}\ new hole-doping regimes may become accessible at fixed compositions by judicious selection of the misfit strain. Given the significant redistribution of charge evident for LNO from our transport measurements, it is feasible that such self-doping effects could allow access to metal-insulator, superconducting, or magnetic transitions without introducing chemical disorder, and in the absence of interface dipoles.

\emph{Acknowledgments.---}%
Majority research at the Univ.\ Arkansas was supported by grants from the DOD-ARO (W911NF-11-1-0200) and partially NSF (DMR-0747808), and at the Univ.\ Miami by an award from Research Corporation. JMR was supported by the Office of Naval Research (ONR N00014-11-1-0664).


\begin{thebibliography}{99}

\bibitem{Mannhart/Schlom:2010}
J. Mannhart and D.G. Schlom, Science {\bf327}, 1607 (2010).

\bibitem{Imada/Fujimori/Tokura:1998}
M. Imada, A. Fujimori, and Y. Tokura, Rev. Mod. Phys. {\bf70}, 1039 (1998).

\bibitem{Pickett:1989}
E.P. Warren, Rev. Mod. Phys. {\bf61}, 433 (1989).

\bibitem{Salamon/Jaime:2001}
M.B. Salamon and M. Jaime, Rev. Mod. Phys. {\bf73}, 583 (2001).

\bibitem{Higuchi/Hwang:2009}
T. Higuchi, Y. Hotta, T. Susaki, A. Fujimori, and H.Y. Hwang, Phys. Rev. B {\bf79}, 075415 (2009).

\bibitem{Hikita/Hwang:2009}
Y. Hikita, M. Nishikawa, T. Yajima, and H.Y. Hwang, Phys. Rev. B {\bf79}, 073101 (2009).

\bibitem{Molegraaf/Ahn/Triscone:2009}
H.J.A. Molegraaf, J. Hoffman, C.A.F. Vaz, S. Gariglio, D. van der Marel \textit{et al.}, Adv. Mater. {\bf21}, 3470 (2009).

\bibitem{Burton/Tsymbal:2011}
J.D. Burton and E.Y. Tsymbal, Phys. Rev. Lett. {\bf106}, 157203 (2011).

\bibitem{Yajima/Hwang:2011}
T. Yajima, Y. Hikita,	and H.Y. Hwang, Nat. Mater.\ {\bf10}, 198 (2011).

\bibitem{Rajeev_et_al:1991}
K.P. Rajeev, G.V. Shivashankar, and A.K. Raychaudhuri, Solid State Commun. {\bf79}, 591 (1991).

\bibitem{Sreedhar/Spalek_et_al:1992}
K. Sreedhar, J.M. Honig, M. Darwin, M. McElfresh, P.M. Shand \textit{et al.}, Phys. Rev. B {\bf46}, 6382 (1992).

\bibitem{Zaanen/Sawatzky/Allen:1984}
J. Zaanen, G.A. Sawatzky, and J.W. Allen, Phys. Rev. Lett. {\bf55}, 418 (1985).

\bibitem{Dobin/Goldman_etal:2003}
A.Y. Dobin, K.R. Nikolaev, I.N. Krivorotov, R.M. Wentzcovitch, E.D. Dahlberg, and A.M. Goldman, Phys. Rev. B {\bf68}, 113408 (2003).

\bibitem{May/Rondinelli:2010}
S.J. May, J.-W. Kim, J.M. Rondinelli, E. Karapetrova, N.A. Spaldin \textit{et al.}, Phys. Rev. B {\bf82}, 014110 (2010).

\bibitem{Boris/Keimer_et_al:2011}
A.V. Boris, Y. Matiks, E. Benckiser, A. Frano, P. Popovich \textit{et al.}, Science {\bf332}, 937 (2011).

\bibitem{Chakhalian/Rondinelli:2011}
J. Chakhalian, J.M. Rondinelli, J. Liu, B.A. Gray, M. Kareev \textit{et al.}, Phys. Rev. Lett. {\bf107}, 116805 (2011).

\bibitem{Han/Millis_etal:2010}
M.J. Han, C.A. Marianetti, and A.J. Millis, Phys. Rev. B {\bf82}, 134408 (2010).

\bibitem{Benckiser/Keimer:2011}
E. Benckiser, M.W. Haverkort, S. Br\"uck, E. Goering, S. Macke \textit{et al.}, Nat. Mater. {\bf10}, 189 (2011).

\bibitem{Stewart/Haule/Chakhalian_etal:2011}
M.K. Stewart, C.-H. Yee, J. Liu, M. Kareev, R.K. Smith \textit{et al.}, Phys. Rev. B {\bf83}, 075125 (2011).

\bibitem{Oullette/Allen_etal:2010}
D.G. Ouellette, S. Lee, J. Son, S. Stemmer, L. Balents \textit{et al.}, Phys. Rev. B {\bf82}, 165112 (2010).

\bibitem{Anisimov/Bukhvalov/Rice:1999}
V.I. Anisimov, D. Bukhvalov, and T.M. Rice, Phys. Rev. B {\bf59}, 7901 (1999).

\bibitem{Chaloupka/Khaliullin:2008}
J. Chaloupka and G. Khaliullin, Phys. Rev. Lett. {\bf100}, 016404 (2008),

\bibitem{Hansmann/Held_et_al:2009}
P. Hansmann, X. Yang, A. Toschi, G. Khaliullin, O.K. Andersen, and K. Held, Phys. Rev. Lett. {\bf103}, 016401 (2009).

\bibitem{Mizokawa/Khomskii/Sawatzky:2000}
T. Mizokawa, D.I. Khomskii, and G.A. Sawatzky, Phys. Rev. B {\bf61}, 11263 (2000).

\bibitem{Kareev/Chakhalian:2011}
M. Kareev, S. Prosandeev, B. Gray, J. Liu, P. Ryan \textit{et al.}, J. Appl. Phys. {\bf109}, 114303 (2011).

\bibitem{Hussey:2003}
N.E. Hussey, Eur. Phys. J. B {\bf31}, 495 (2003).

\bibitem{Hussey:2008}
N.E. Hussey, J. Phys. Condens. Mater. {\bf20}, 123201 (2008).

\bibitem{Ioffe/Millis:1998}
L.B. Ioffe and A.J. Millis, Phys. Rev. B {\bf58}, 11631 (1998).

\bibitem{Konstantinovi:2000}
Z. Konstantinovi\'{c}, Z.Z. Li, and H. Raffy, Phys. Rev. B {\bf62}, R11989 (2000).

\bibitem{Ando/Murayama:1999}
Y. Ando and T. Murayama, Phys. Rev. B {\bf60}, R6991 (1999).

\bibitem{Eguchi/Shin_et_al:2009}
R. Eguchi, A. Chainani, M. Taguchi, M. Matsunami, Y. Ishida \textit{et al.}, Phys. Rev. B {\bf79}, 115122 (2009).

\bibitem{footnote1}
We follow the procedure reported in Ref.\cite{May/Rondinelli:2010}, but note a few differences: we omit structural distortions due to NiO$_6$ rotations. At the bulk level, the electronic structure that results when they are included is nearly completely accounted for by zone-folding arguments. We anticipate this to be also true in the ultra-thin films and therefore can be neglected. We also sample the BZ with a denser $11\times 11\times 11$ $k$-point mesh and perform integrations using a Gaussian smearing of 5~meV.

\bibitem{Jepsen/Andersen:1995}
O. Jepsen and O.K. Andersen, Zeitschrift f{\"u}r Physik, {\bf97} (1995).

\bibitem{Korotin/Sawatzky:1998}
M.A. Korotin, V.I. Anisimov, D.I. Khomskii, and G.A. Sawatzky, Phys. Rev. Lett. {\bf80}, 4305 (1998).

\bibitem{Khomskii:1997}
D.I. Khomskii, Lith. J. Phys. {\bf 37}, 65 (1997).

\bibitem{footnote2}
Both changes in stoichiometry and chemistry are prohibited in calculations.

\bibitem{footnoteYAO} 
That the YAO film is an outlier in its TEP magnitude may be connected to domain formation due to anisotropy of the in-plane strain (YAlO$_3$ is orthorhombic) or thermal expansion [see, \textit{e.g.} O. Chaix-Pluchery, B. Chenevier, and J.J. Robles, Appl. Phys. Lett. {\bf 86}, 251911 (2005)].

\bibitem{footnote3}
From Boltzmann theory, $S = (\pi ^2/3e)k_{B}^2T(\partial \protect \qopname  \relax o{ln}\sigma /\partial \varepsilon )|_{\varepsilon _F}$;
see, e.g., F.J. Blatt {\protect \it et al.}, {\protect \it Thermoelectric Power of Metals} (Plenum Press, 1976).

\bibitem{Obertelli:1992}
S.D. Obertelli, J.R. Cooper, and J.L. Tallon, Phys. Rev. B {\bf46}, 14928 (1992).

\bibitem{Honma/Hor:2008}
T. Honma and P.H. Hor, Phys. Rev. B {\bf77}, 184520 (2008).

\bibitem{Greene:98} 
N. Gayathriy, A.K. Raychaudhuri, X.Q. Xu, J.L. Peng, and R.L. Greene, J. Phys.: Condens. Matter {\bf 10}, 1323 (1998).

\bibitem{Khomskii/Sawatzky:1997}
D.I. Khomskii and G.A. Sawatzky, Solid State Commun. {\bf 102}, 87 (1997).

\end{thebibliography}
\end{document}